\documentclass[12pt,preprint]{aastex}

\def \Lo {$\,$L$_{\odot}$}
\def \Mo {$\,$M$_{\odot}$}
\def \kms {$\,$km s$^{-1}$}

\def \mjyb {$\,$mJy$\,$beam$^{-1}$}

\def \asec{{\em $.\!\!^{\prime\prime}$}}
\def \arcsec{{\em $^{\prime\prime}$}}
\def \arcmin{{\em $^\prime$}}
\def \arcdeg{{\em $^o$}}

\def\h{\hbox{$^{\rm h}$}}
\def\m{\hbox{$^{\rm m}$}}

\def\fsecs{\hbox{$.\!\!^{\rm s}$}}

\shorttitle{H$_2$CO Galactic Emission. III}
\shortauthors{Araya et al.}

\begin{document}

\title{A Search for H$_2$CO 6$\,$cm Emission toward Young Stellar Objects III: 
VLA Observations}


\author{E. D. Araya \altaffilmark{1,}\altaffilmark{2,}\altaffilmark{3},
P. Hofner\altaffilmark{2,}\altaffilmark{3},
W. M. Goss\altaffilmark{2},
H. Linz\altaffilmark{4},
S. Kurtz\altaffilmark{5}
\&
L. Olmi\altaffilmark{6,7} 
}
\altaffiltext{1}{Department of Physics and Astronomy, 
MSC07 4220, University of New Mexico, Albuquerque, NM 87131}
\altaffiltext{2}{National Radio Astronomy Observatory, P.O. 
Box 0, Socorro, NM 87801}
\altaffiltext{3}{New Mexico Institute of Mining and Technology, 
Physics Department, 801 Leroy Place, Socorro, NM 87801}
\altaffiltext{4}{Max--Planck--Institut f\"ur Astronomie, K\"onigstuhl 17,
D--69117 Heidelberg, Germany.}
\altaffiltext{5}{Centro de Radioastronom\'{\i}a y Astrof\'{\i}sica,
Universidad Nacional Aut\'onoma de M\'exico,
Apdo. Postal 3-72, 58089, Morelia, Michoac\'an, Mexico.}
\altaffiltext{6}{University of Puerto Rico at Rio Piedras, Physics
Department, P.O. Box 23343, San Juan, PR 00931.}
\altaffiltext{7}{Istituto di Radioastronomia, INAF, Sezione di Firenze,
Largo Enrico Fermi 5, I-50125 Florence, Italy.}

\begin{abstract}

We report the results of our third survey for formaldehyde
(H$_2$CO) 6$\,$cm maser emission in the Galaxy.
Using the Very Large Array, we detected
two new H$_2$CO maser sources (G23.01$-$0.41 and G25.83$-$0.18),
thus increasing the sample of known H$_2$CO maser regions
in the Galaxy to seven. We review the characteristics of the 
G23.01$-$0.41 and G25.83$-$0.18 star forming regions.
The H$_2$CO masers in G23.01$-$0.41 
and G25.83$-$0.18 share several properties with the other
known H$_2$CO masers, in particular, emission from
rich maser environments and close 
proximity to very young massive stellar 
objects. 

\end{abstract}

\keywords{ISM: molecules --- masers --- radio lines: ISM}

~
\clearpage

\section{Introduction} 

The first formaldehyde (H$_2$CO) 6$\,$cm maser was discovered in 1974 
toward NGC$\,$7538 (Downes \& Wilson 1974; 
Forster et al. 1980); in the following $\sim 28\,$years, H$_2$CO masers
were detected only toward two other regions in the Galaxy:
Sgr B2 (Whiteoak \& Gardner 1983) and G29.96$-$0.02
(Pratap et al. 1994). 
This low number of H$_2$CO maser regions 
was unexpected (e.g., Forster et al. 1985; Gardner et al. 1986) 
given the detection of many H$_2$CO maser spots 
in a single source (Sgr B2, Whiteoak \& Gardner 1983;
Mehringer et al. 1994), the widespread distribution of
formaldehyde molecules as exemplified by 
Galactic H$_2$CO 6$\,$cm absorption
(e.g., Watson et al. 2003; Araya et al. 2002; 
Downes et al. 1980; Dieter 1973),
and the apparently common astrophysical conditions
needed for the maser excitation if the masers
were pumped by radio continuum radiation (see Pratap et al. 1992).

The idea that H$_2$CO masers may be pumped by 
background radio continuum was initially proposed by  
Boland \& de Jong (1981) to explain the maser in NGC$\,$7538 IRS1. 
However, the low detection rate of new H$_2$CO masers in dedicated 
surveys (Mehringer et al. 1995; Forster et al. 1985),
the non-detection of maser emission from the
H$_2$CO 2$\,$cm transition (e.g., Hoffman et al. 2003) and
low emission measure (or non-detection) of radio continuum 
sources near several of the known H$_2$CO masers, 
indicate that the pumping mechanism in most cases cannot be
due to radio continuum excitation (e.g., see 
Araya et al. 2007c for the case of IRAS$\,$18566+0408).
The low detection rate of H$_2$CO masers led several 
authors to speculate that formaldehyde masers are 
rare because specific and/or short-lived physical conditions
may be needed for the excitation (e.g., Forster et al. 1985; 
Mehringer et al. 1995; Araya et al. 2007a).

In an effort to understand the H$_2$CO maser phenomenon and
its place during the formation of massive stars, we
conducted two surveys for H$_2$CO masers using
Arecibo, the Green Bank Telescope, and the 
VLA\footnote{The Very Large
Array (VLA) is operated by the National Radio Astronomy Observatory
(NRAO), a facility of the National Science Foundation operated
under cooperative agreement by Associated Universities, Inc.} 
that resulted in the detection of two new Galactic H$_2$CO maser
regions (IRAS$\,$18566+0408, with emission first 
detected by Araya et al. 2004 and confirmed to be a maser by 
Araya et al 2005;  and G23.71-0.20 with details in Araya et al. 
2006, from a survey later reported by Araya et al. 2007b).
In this article we present the results of our third
survey for H$_2$CO masers.

\section{Observations}\label{obs}

\subsection{VLA Survey}

The initial observations were conducted in 2006 June with the VLA
during the BnA $\rightarrow$ B reconfiguration 
($\theta_{syn}\sim 1.5$\arcsec).
The details of the observations are reported in Table~1. 
A total of 14 pointing positions were observed toward
massive star forming regions. In some cases, there 
were multiple radio continuum objects within the primary beam of
a given position.
Nine of the targets were selected based on their 
GBT and Arecibo spectra (Araya et al. 2002; Watson et al. 
2003; Sewi{\l}o et al. 2004; Araya et al. 2007b) that
showed profiles suggesting
H$_2$CO emission blended with
absorption. In addition, given that H$_2$CO 6$\,$cm masers appear to 
originate mostly from regions that also harbor H$_2$O and Class II
CH$_3$OH masers (e.g., Araya et al. 2005), 
we complemented our sample with 5 targets from the methanol
and water maser catalog of Szymczak et al. (2005). The targets from the 
Szymczak et al. (2005) sample were observed $\sim$40$\,$min
on-source. In the case of the targets that were selected based on 
single dish H$_2$CO spectra, the time on-source ranged between
10 to 100$\,$min depending on the expected intensity
of the H$_2$CO emission line candidates.
In Table~2 we list the observed targets, phase tracking center,
center bandpass velocity, secondary (phase) calibrator used, 
and single-channel rms of the final data cubes.
The data reduction and imaging were conducted using the 
NRAO package AIPS following the standard procedure 
for spectral line observations.

\subsection{High Spectral Resolution Observations}

Follow-up H$_2$CO 6$\,$cm observations of G23.01$-$0.41 and
G25.83$-$0.18 were conducted on 2007 February 16 with
the VLA in the D configuration. The goals of the observations
were to confirm the H$_2$CO 6$\,$cm emission detected
in our survey ($\S3$), and to study the 
H$_2$CO maser line profiles with higher spectral resolution
(1.83$\,$kHz, 0.11\kms). The VLA correlator
was used in the 1 IF mode with a bandwidth of 0.78$\,$MHz
(48\kms) and 512 channels. The central bandpass velocity
was set to 70 and 87\kms~for the G23.01$-$0.41
and G25.83$-$0.18 observations, respectively.
The phase tracking center of the G23.01$-$0.41 observations
was RA = 18\h34\m40\fsecs30, Decl. = $-$09\arcdeg00\arcmin00\farcs0
(J2000); the phase tracking center of the G25.83$-$0.18
observations was RA = 18\h39\m03\fsecs60, 
Decl. = $-$06\arcdeg24\arcmin50\farcs0 (J2000).
The quasars 3C286 and J1822$-$096 were used as primary
and secondary calibrators, respectively. We assumed a flux density of 
7.52$\,$Jy for 3C286, and measured a flux density of
2.29$\,$Jy for J1822$-$096. In addition, 3C48 was
observed to check the flux density calibration.
Using 3C286 as primary calibrator to bootstrap 
the flux density of 3C48, we measured a flux density of 
5.60$\,$Jy for 3C48, which 
agrees within 3$\%$ with the expected value of
5.47$\,$Jy. The calibration and imaging were conducted
using the NRAO package AIPS.

\section{Results}\label{results}

Of the 14 pointing positions observed with the VLA
(and a total of 18 targets, see Table~2) we detected H$_2$CO
6$\,$cm emission toward two sources: 
G23.01$-$0.41 and G25.83$-$0.18 (Figure~1). In both cases
the emission was spatially unresolved 
(deconvolved size $<$ 0.5\arcsec) implying
brightness temperatures greater than 
$10^4\,$K, which indicates maser emission.
Thus, G23.01$-$0.41 and G25.83$-$0.18 are the 
sixth and seventh regions in the Galaxy where
H$_2$CO 6$\,$cm masers have been detected. 

Figure~1 shows the low velocity resolution spectra
and peak channel images 
of the two new maser regions. The G23.01$-$0.41 line
is spectrally unresolved whereas the H$_2$CO 
emission in G25.83$-$0.18 has a double peaked profile. 
Figure~2 shows the line profiles of the high spectral
resolution (VLA-D, \S2.2) observations. H$_2$CO
emission was confirmed in both cases.
In Table~3 we list the line parameters of the 
H$_2$CO 6$\,$cm masers from both observing
periods.

We detected radio continuum emission as well
as H$_2$CO 6$\,$cm absorption toward several
of the targets in the sample; 
a discussion of these
data will be the topic of a future paper. 
Here we focus on the H$_2$CO 6$\,$cm
maser emission, and discuss 
the H$_2$CO 6$\,$cm absorption only in the
case of G25.83$-$0.18 since the H$_2$CO
maser originates at the (projected)
spatial center of a molecular clump traced by H$_2$CO 
absorption ($\S4.2$).

\section{Discussion}\label{discussion}

\subsection{G23.01$-$0.41}

\subsubsection{H$_2$CO 6$\,$cm Maser Emission}

H$_2$CO 6$\,$cm maser emission was detected
toward G23.01$-$0.41 with the VLA in
the BnA $\rightarrow$ B reconfiguration (Figure~1),
and confirmed with VLA-D observations conducted
approximately 8 months later (\S2.2,
Figure~2). The peak position, LSR velocity, and line
width measurements are consistent in both runs
(Table~3). If the VLA-D spectrum is smoothed 
to the channel width of the lower spectral resolution 
observations (VLA BnA$\rightarrow$B, Table~3), the peak
intensities are consistent within 2$\sigma$, revealing
no apparent variability of the line in a 
$\sim 8$ month period.

As mentioned in $\S3$, the H$_2$CO 6$\,$cm
maser in G23.01$-$0.41 was not spectrally
resolved in the VLA BnA$\rightarrow$B 
observations (line width $<$ 1.0\kms, Figure~1).
In the high spectral resolution observations
(Figure~2) the line was barely resolved, 
and appears to be the superposition of
two components. The separation  
of the two possible overlapping lines is 
$\la$0.4\kms.

\subsubsection{Infrared Environment}

Far-IR emission is found at the location of the 
H$_2$CO maser as shown by IRAS 
data\footnote{IRSA-IRAS Sky Survey Atlas,
http://irsa.ipac.caltech.edu/Missions/iras.html}. 
However, the far-IR emission at the H$_2$CO 
maser location is overlapped with several nearby
sources. MSX 21.4$\mu$m data toward G23.01--0.41 is also 
affected by confusion due to blended emission
with a source $\sim 20$\arcsec~east of the
H$_2$CO maser position. At 70$\mu$m, MIPSGAL Spitzer data show 
that the maser is coincident with a compact (FWHM $<$ 30\arcsec)
and strong (saturated) FIR source. Since the emission is
saturated, a reliable measurement of the 70$\,\mu$m size and
position cannot be obtained; nevertheless, the peak position
appears to be within $\sim 3$\arcsec~of the H$_2$CO maser, i.e., 
certainly within the telescope angular resolution at 70$\,\mu$m
($\sim 20$\arcsec; Rieke et al. 2004). At the 24$\,\mu$m 
Spitzer MIPS band, the IR source coincident with the H$_2$CO maser
is substantially less affected by saturation and the center 
position can be more reliably determined; the 
24$\,\mu$m source and the H$_2$CO maser are less that 
0.3\arcsec~apart.

The IR source coincident with the H$_2$CO maser
is detected in mid/near-IR observations from 
the Spitzer/IRAC GLIMPSE survey (Benjamin et al. 2003). 
Figure~3 (upper panel) shows a three color image 
(3.6$\mu$m blue, 4.5$\mu$m green, 8.0$\mu$m red) of the G23.01$-$0.41
region. The Spitzer source at the H$_2$CO maser location
shows strong 4.5$\mu$m excess indicative of 
shocked gas (e.g., Smith et al. 2006). There is also extended 
4.5$\mu$m excess emission west 
of the H$_2$CO maser position (Figure~3, inset upper panel).

Testi et al. (1998) detected a near infrared (NIR)
source toward G23.01$-$0.41, that was also detected by
2MASS. The 2MASS source is $\sim 4$\arcsec~displaced 
from the H$_2$CO maser position. Given this large angular
offset, it is unlikely that the NIR source is 
responsible for the excitation of the masers
in G23.01$-$0.41 ($\S 4.1.4$). 

\vspace{-0.5cm}

\subsubsection{Radio Continuum and Thermal Molecular Lines}

We detected no 6$\,$cm radio continuum toward the 
maser position to a 5$\sigma$ level
of 2.4\mjyb~($\theta_{syn} = 1$\asec$6\times1$\asec1).
Using the VLA in the C configuration ($\theta_{sys} \sim 1$\arcsec), 
Codella et al. (1997) report an upper limit of the
1.3$\,$cm radio continuum of 0.4\mjyb~(3$\sigma$).

Based on our VLA-D observations, 
the nearest 6$\,$cm radio continuum source 
is approximately 50\arcsec~north of the maser
position. The radio continuum source (a compact
H{~\small II} region that will be discussed 
in a future paper) has associated 
H$_2$CO absorption at approximately the same velocity 
as the H$_2$CO maser, indicating that the H$_2$CO 
maser belongs to an extended
molecular cloud that shows clear evidence for the presence of
more evolved massive stars.
The LSR velocity of the H$_2$CO absorption
(75.0\kms) implies two possible kinematic 
distances: 4.8 or 10.8$\,$kpc. Codella et al. (1997)
as well as Caswell \& Haynes
(1983) preferred the far kinematic distance\footnote{Caswell \& Haynes
(1983; see also Caswell et al. 1995b, Anglada et al. 1996,
Testi et al. 1998, Forster \& Caswell 1999) 
report a far kinematic distance of 12.8$\,$kpc instead
of our value of 10.8$\,$kpc because they assumed a 
Sun -- Galactic Center distance of 10.0$\,$kpc, whereas
we assume 8.5$\,$kpc (Brand \& Blitz 1993).}; 
however the distance ambiguity has not been 
resolved (Forster \& Caswell 1999). For example, Harju et al. (1998)
list the near kinematic distance for the source
(see also Scoville et al. 1987, Pestalozzi et al. 2005). 

A number of thermal molecular lines have been
detected toward G23.01$-$0.41. 
Caswell et al. (2000) detected (quasi-)thermal emission of the
CH$_3$OH lines at 107.0$\,$GHz\footnote{A maser component
was also found superimposed on the thermal profile of the 
CH$_3$OH 107.0$\,$GHz line.} and 156.6$\,$GHz (see also
Slysh et al. 1999). Harju et al. (1998)
detected broad (line width $>$ 20\kms) 
SiO J=2--1 and J=3--2 emission with the SEST telescope. 
CO and $^{13}$CO emission was also detected 
in single dish surveys
(Scoville et al. 1987; Jackson et al. 2006).

Interferometric (Nobeyama Millimeter Array and
Plateau de Bure Interferometer) observations of $^{12}$CO, 
$^{13}$CO, and C$^{18}$O have been recently
reported by Furuya et al. (2008) with an angular
resolution $\sim$6\arcsec.
Their $^{12}$CO spectrum shows high velocity gas,
in particular prominent red-wing
emission that traces a massive ($>50$\Mo) molecular outflow,
possibly located close to the plane of the sky.

High density molecular gas in the
region is evident from detection of CS ($J=1-0$)
and NH$_3$ (1,1) lines with the MIT Haystack 37$\,$m telescope
(Anglada et al. 1996).
Single dish NH$_3$ (1,1), (2,2), and (3,3) 
observations were also conducted by Codella et al. (1997) 
with the Medicina 32$\,$m telescope. 
Codella et al. (1997) derived a rotational temperature of 13$\,$K. 

High angular resolution NH$_3$ observations were
conducted by Codella et al. (1997) with the 
VLA ($\theta_{syn} \sim 1$\arcsec).
The NH$_3$ (3,3) data show a compact ($<10$\arcsec) molecular core
that has a SE--NW elongation (Figure~3, lower panel). 
Codella et al. (1997) derived the following
parameters for the ammonia core: deconvolved angular diameter = 3\asec3,
T$_k = 58\,$K, M$_{\mathrm{vir}} = 886$\Mo, and 
$n_{\mathrm{H_2}} = 6.9 \times 10^6\,$cm$^{-3}$.
The H$_2$CO maser is coincident with the NH$_3$ (3,3) 
peak emission (Figure~3, lower panel). 

Furuya et al. (2008) recently conducted a detailed
multi-wavelength study of G23.01$-$0.41 at high
angular resolution ($<$ 10\arcsec) 
using CO (see above), HNCO, and CH$_3$CN. 
Their observations confirm the presence of a
hot molecular core characterized
by a CH$_3$CN rotation temperature of $\sim 120\,$K
and a core mass of $\sim 380$\Mo.
The velocity distribution of the CH$_3$CN 
gas is consistent with rotation of a molecular core
oriented almost perpendicular to the outflow direction. 
The elongation and velocity gradient of the 
CH$_3$CN core is parallel to the elongation of
the NH$_3$ emission shown in Figure~3.

\subsubsection{Other Astrophysical Masers}

A variety of masers have been detected 
toward G23.01$-$0.41 
in single dish surveys.
Methanol masers have been found at
6.7$\,$GHz (Menten 1991;
Caswell et al. 1995b; Szymczak et al. 2002;
see also catalogs by Pestalozzi et al. 2005 
and Xu et al. 2003), 12$\,$GHz
(MacLeod et al. 1993; Caswell et al. 1995a; 
B{\l}aszkiewicz \& Kus 2004), 107.0$\,$GHz
(Caswell et al. 2000), 44$\,$GHz 
(Slysh et al. 1994), and 95$\,$GHz
(Val'tts et al. 2000).
H$_2$O 22$\,$GHz and OH ($^2\Pi_{3/2}~J=3/2$ ground 
state) masers have also been reported
(Szymczak et al. 2005; Szymczak \& G\'erard 2004; 
Caswell \& Haynes 1983). 

High angular resolution observations of the CH$_3$OH 6.7$\,$GHz masers
have been conducted (Caswell, unpublished data; see 
Caswell et al. 2000); the position of the CH$_3$OH 
maser is coincident with that of the H$_2$CO maser
within 1\arcsec~rms.
Forster \& Caswell (1989, see also Forster \& Caswell 1999)
conducted VLA observations of H$_2$O 22$\,$GHz 
($\theta_{syn} = 3$\asec$9\times1$\asec8)
and OH 1665$\,$MHz 
($\theta_{syn} = 6$\asec$5\times1$\asec0)
masers in G23.01$-$0.41. 
The H$_2$CO maser is located within $2$\arcsec~of 
the OH and H$_2$O maser positions (see Figure~3 lower panel).

\subsubsection{The Nature of the G23.01$-$0.41 Star Forming Region}

The abundant available multi-wavelength data
shows that G23.01$-$0.41 is an active site
of massive star formation. The H$_2$CO maser
is located at the peak of a molecular core
traced by NH$_3$ (3,3) and CH$_3$CN. The hot 
molecular core also harbors a variety of 
molecular masers. Given the highly confused
far IR environment, it is not possible to 
reliably measure the bolometric luminosity of the massive
stellar object pinpointed by the H$_2$CO
maser (Figure~3); however, based on the available mid and
far IR data, the upper limit of the luminosity is 
$\sim 10^6$\Lo. As in the case of the infrared source
associated with the H$_2$CO maser in IRAS$\,18566+0408$
(Araya et al. 2007c), the H$_2$CO maser in G23.01$-$0.41
is found toward a source with 4.5$\mu$m
infrared excess, which likely indicates shocked
gas in an outflow. Considering the presence
of a hot molecular core at the location of the 
H$_2$CO maser, the detection of several other maser
species, evidence for outflow and shocked gas based on
CO, SiO and 4.5$\mu$m infrared excess emission, absence of radio
continuum, and evidence for rotation of the molecular
core, the H$_2$CO maser appears to pinpoint the 
location of a very young massive stellar object in a
evolutionary stage prior to the formation of a radio bright
ultra-compact H{~\small II} region. The central object 
may still be undergoing accretion; further observations 
should be made to clarify this point.

\subsection{G25.83$-$0.18}

\subsubsection{H$_2$CO 6$\,$cm Maser Emission}

An the case of G23.01$-$0.41 ($\S4.1$), maser emission
in G25.83$-$0.18 was first detected in the VLA
survey ($\S2.1$) and then confirmed
with higher spectral resolution observations 
(VLA-D, $\S2.2$) $\sim 8\,$months later. The H$_2$CO maser
shows a double peaked line profile (Figures~1 and 2).
The separation between the two peak components is 1.5\kms.
No significant difference in the maser flux density
was found between the two epochs.

As reported in Table~3, the line profile of the high spectral resolution
observations is well fit by the superposition of three 
Gaussian profiles; two broad components (0.7 and 0.9\kms~FWHM)
and a narrow component (FWHM = 0.29\kms). Whether the 
line profile is composed of only two non-Gaussian maser lines
or three (or more) components is unclear.
For example, the profile could be due to an asymmetric
line (the blue-shifted component) overlapped with a Gaussian 
line (the red-shifted component, see Figure~2).

\subsubsection{Infrared Environment}

The H$_2$CO maser is at the center of an infrared dark 
cloud as revealed by 8.0$\mu$m \linebreak Spitzer/GLIMPSE 
observations (see Figure~4). The H$_2$CO
maser is located between a source with strong
4.5$\mu$m excess emission (green in Figure~4, upper
panel inset) and a source
brighter at 8$\mu$m (red in Figure~4). The peak of the 
4.5$\mu$m source is offset ($\sim 3$\arcsec) 
from the location of the H$_2$CO maser. No infrared
emission was detected with MSX toward the position of the
H$_2$CO maser; no 2MASS source was found coincident with the 
H$_2$CO maser either. 

Data from Spitzer MIPSGAL at 70$\mu$m reveal a strong
($>100\,$Jy) far IR source whose peak is within 
$\sim$1\arcsec~from the position of the H$_2$CO maser. 
At low level, the 70$\mu$m emission is extended and comprises
a neighboring infrared source; however, the core emission
is compact (FWHM$\,<\,$25\arcsec), i.e., close to the theoretical 
telescope diffraction limit. The IR source coincident with 
the H$_2$CO maser is also detected in the 24$\mu$m MIPS band.

\subsubsection{Radio Continuum and Thermal Molecular Lines}

We detected no radio continuum toward the position of the
H$_2$CO maser to a level of 2.8\mjyb~(5$\sigma$) in the
VLA BnA $\rightarrow$ B observations
($\theta_{syn} = 1$\asec$6\times 1$\asec$1$); the 5$\sigma$ upper
limit set by the VLA-D data is 5\mjyb~(see Figure~4, upper
panel). The nearest radio continuum source to the H$_2$CO maser
is $\sim$2\arcmin~to the west (G25.80$-$0.16;
Figure~4, upper panel) which was also detected in the continuum
at 8.64 and 6.67$\,$GHz by Walsh et al. (1998).

The H$_2$CO maser is coincident with a mm and sub-mm core.
Walsh et al. (2003) detected 
compact 450$\mu$m and 850$\mu$m emission; the H$_2$CO
maser is located within 4\arcsec~of the peak position
of the sub-mm source (the JCMT beam is
approximately 8\arcsec~and 15\arcsec~at 450$\mu$m
and 850$\mu$m, respectively). Hill et al. (2005) conducted
SEST/SIMBA observations of G25.83$-$0.18 and detected
1.2$\,$mm emission ($S_\nu = 5.4\,$Jy); the FWHM of the mm source is 
60\arcsec. Based on the mm detection, Hill et al. (2005) report 
a mass of $2.8 \times 10^3$\Mo~or
$8.8 \times 10^3$\Mo~depending on the kinematic distance
(see below). The H$_2$CO maser is coincident with 
the 1.2$\,$mm peak within $\sim 1$\arcsec.

We detected H$_2$CO 6$\,$cm absorption in the region
(Figure~4, middle and lower panels, VLA-D observations). 
The main H$_2$CO absorption clump is coincident with the infrared dark
cloud, and shows a shell-like brightness distribution
(Figure~4, middle panel)\footnote{The
shell-like distribution of 
H$_2$CO absorption (Figure~4, middle panel) is not due 
to overlapping H$_2$CO maser emission and absorption at the 
central position; the measured velocity of the 
H$_2$CO maser is not in the velocity
range used to obtain the H$_2$CO absorption image.}.
The H$_2$CO maser is located at the projected 
spatial center of the shell and near the radial 
velocity edge of the H$_2$CO absorption line (Figure~4, lower panel).  
The shell brightness distribution 
could be due to: 1.) H$_2$CO in gas phase 
is less abundant in the inner regions of the shell, for example,
due to chemical gradients, smaller
total molecular density, or depletion (see Young et al.
2004 for the case of pre-protostellar cores); and/or
2.) the excitation conditions for H$_2$CO absorption are
less favorable in the inner regions of the clump (e.g., Zhou et al.
1990 explained the detection of an H$_2$CO shell-like structure
in B335 as a consequence of higher molecular density at the core 
that quenches the anomalous absorption). Observations
of other H$_2$CO transitions as well as other mid and high
density tracers are needed to fully investigate the nature
of the shell-like H$_2$CO absorption source. Nevertheless, the recent
high angular resolution detection of NH$_3$ at the position 
of the H$_2$CO depression (Longmore et al. 2007, see below) 
suggests that the H$_2$CO shell-like structure is due to
high density ($>10^5\,$cm$^{-3}$) molecular gas in the center 
of the molecular core that quenches H$_2$CO anomalous absorption.

We also detected H$_2$CO absorption close to the
radio continuum source (compare Figure~4 upper and middle panels).
The similar velocity of the H$_2$CO absorption gas 
associated with the continuum source and with the infrared 
dark cloud indicates that both are part of the same star
forming complex. Assuming that the LSR velocity of the
H$_2$CO absorption line traces the systemic velocity of the 
cloud, the two possible kinematic distances to this massive
star forming region are 5.6 and 9.7$\,$kpc. 
Since the H$_2$CO maser is associated
with an infrared cloud seen in absorption against the 
mid-IR galactic background (Figure~4), 
it is likely that the region is located 
at the near kinematic position. 

G25.83$-$0.18 has been detected in CH$_3$CN, HCO$^+$,
and H$^{13}$CO$^+$ with the Mopra telescope by
Purcell et al. (2006). Based on the CH$_3$CN data, 
they derived a rotation temperature
of $\sim 50\,$K, and found evidence for infalling motion 
based on HCO$^+$ data. In addition, Longmore et al. (2007)
detected an optically thick NH$_3$ core at the position of the H$_2$CO 
maser from ATCA observations. The line width 
of the NH$_3$ (1,1), (2,2), (4,4), and (5,5) transitions
range between 6 and 30\kms, indicating the possible presence
of a molecular outflow partially traced by NH$_3$.

\subsubsection{Other Astrophysical Masers}

Walsh et al. (1998) conducted ATCA observations of 
6.7$\,$GHz CH$_3$OH masers in \linebreak G25.83$-$0.18; 
five maser components were found with LSR velocities between
of 90.7 and 99.0\kms. The H$_2$CO maser
is located $\sim 2$\arcsec~south of the CH$_3$OH maser
clump\footnote{The absolute position accuracy of the 
CH$_3$OH masers is $\sim 1$\arcsec, Walsh et al. (1998).}
(Figure~4, upper panel, inset).

Single dish observations of the Class II CH$_3$OH
6.7 and 12$\,$GHz masers were conducted by
B{\l}aszkiewicz \& Kus (2004) with the Toru\'n telescope.
The peak LSR velocity of the 6.7 and 12$\,$GHz masers were 
91.3 and 90.7\kms, i.e., coincident in velocity with the H$_2$CO maser.

Ellingsen (2005) conducted observations with the Mopra
Telescope of the 95.1 GHz Class I CH$_3$OH maser
transition, and detected several maser lines (peak maser 
emission at 90.2\kms) overlapped with a broad component. 
The broad line (4.3\kms~line width) could be
due to thermal emission given that 
its peak velocity (94.2\kms) coincides with the H$_2$CO
absorption peak velocity (see $\S4.2.3$).

Other maser species have also been detected in 
single dish surveys. Szymczak \& G\'erard (2004) 
detected a OH 1667$\,$MHz maser line with a peak 
LSR velocity of 92.9\kms; 
their figure A.1 also shows detection of a possible
high velocity ($\sim 120$\kms) OH 1612$\,$MHz line.
Szymczak et al. (2005) conducted H$_2$O 22$\,$GHz
observations with the 100$\,$m Effelsberg telescope
and detected several H$_2$O maser features within a
velocity range of $\sim 50$\kms, centered at
94.7\kms.

\subsubsection{The Nature of the G25.83$-$0.18 Star Forming Region}

The coincidence of the H$_2$CO maser with
other maser species (in particular with Class II CH$_3$OH masers), 
with an infrared dark cloud and molecular core,
the absence of radio continuum emission, the detection
of a mm, sub-mm and far-IR source, and the presence 
of 4.5$\mu$m excess emission toward the center of the 
infrared dark cloud, imply that
G25.83$-$0.18 is a very young region of massive star
formation in an evolutionary stage prior to the 
ultra-compact H{~\small II} region phase. The precise location and luminosity
of the protostar responsible for the excitation of the different maser 
species is unclear;
higher sensitivity continuum observations are needed to
reveal the position of the exciting source. 
G25.83$-$0.18, as well as G23.01$-$0.41, may be classified
as Group 2 cores following the Longmore et al. (2007) nomenclature, 
i.e., warm NH$_3$ cores associated with CH$_3$OH masers but no
detectable radio continuum.

\subsection{G23.01$-$0.41 and G25.83$-$0.18 with respect to
the other known H$_2$CO Maser Regions}

Including the two new masers reported in this work, 
H$_2$CO 6$\,$cm masers have been detected toward seven regions in 
the Galaxy, and in a total of 15 maser `spots'
(at 1\arcsec~resolution). 
The H$_2$CO masers in G23.01$-$0.41 and G25.83$-$0.18 share
similar characteristics with most of the other known
H$_2$CO maser regions, in particular:
(1.) H$_2$CO masers are found in regions that harbor 
a variety of other molecular masers (e.g., Hoffman et al. 2003;
Mehringer et al. 1994), (2.) the velocity difference
of the H$_2$CO masers with respect to the systemic
velocity of the clouds is typically less than 6\kms (e.g., 
Pratap et al. 1994; Araya et al. 2004), 
which suggests that H$_2$CO masers do not originate in 
high velocity outflows, (3.) the flux density of the known
masers is less than $\sim$2$\,$Jy (e.g., Hoffman et al. 2007), 
(4.) all known H$_2$CO masers have been detected in regions 
of massive star formation (Araya et al. in prep.), 
(5.) excluding Sgr B2, most of the 
H$_2$CO masers show double peaked profiles with separations
smaller than 3\kms~(e.g., NGC$\,$7538 IRS1, Forster et al. 1985; 
IRAS$\,$18566+0408, Araya et al. 2007d; see also review by 
Araya et al. 2007a), (6.) even though bright radio continuum sources may
be found in the same star forming complexes,  
most H$_2$CO masers are located toward
sources characterized by weak (or no) compact radio continuum emission, 
typically undetected at a few mJy sensitivity levels
(e.g., Araya et al. 2005; this work),
(7) excluding some of the masers in Sgr B2, the known
H$_2$CO masers appear to be associated with very young
massive stellar objects (in an evolutionary phase prior
to the formation of radio-bright ultra-compact H{~\small II} regions) that have
strong far IR emission and molecular core counterparts
(e.g., $\S 4.1$ and 4.2).

Many H$_2$CO maser regions show evidence of outflows/jets and shocked gas
based on 4.5$\mu$m excess emission
from Spitzer/IRAC data (see, for example, Araya et al. 2007c
in the case of IRAS$\,$18566+0408). Further evidence comes from
molecular data such as SiO, H$_2$, and H$_2$S 
(e.g., Zhang et al. 2007; Beuther et al. 2007b; Kraus et al. 2006;
Gibb et al. 2004; Maxia et al. 2001; Harju et al. 1998),
and radio continuum observations (e.g., Araya et al. 2007c). However,
the H$_2$CO masers may not be directly associated with
the shocked material (e.g., note the offset between the 
4.5$\mu$m excess source and the H$_2$CO maser in 
G25.83$-$0.18; Figure~4, upper panel).

\vspace*{-0.5cm}

\section{Summary}

We report the results of our third survey for H$_2$CO 6$\,$cm
masers toward massive star forming regions in the Galaxy. 
The observations
were conducted with the VLA toward 14 pointing positions, 
and resulted in the detection of two
new H$_2$CO maser regions: G23.01$-$0.41 and
G25.83$-$0.18. Including the new detections, H$_2$CO masers have been
found toward a total of seven star forming regions in the Galaxy, 
four of them detected in our series of surveys (Araya et al. 2004,
2007b, and this work). 

The H$_2$CO maser in G23.01$-$0.41 is coincident with 
the center of a hot molecular core that shows evidence of
a rotating torus perpendicular to a molecular outflow 
(Furuya et al. 2008). Excess in the 4.5$\,\mu$m band from 
GLIMPSE Spitzer/IRAC observations reveals shocked gas in the
region, possibly tracing the outflow. Active massive star formation is also 
evident from the detection of a number of molecular maser lines,
including Class II CH$_3$OH masers.

In the case of G25.83$-$0.18, the maser is located at the
center of an infrared dark cloud that harbors a clump
of CH$_3$OH 6.7$\,$GHz masers, a high density molecular
core traced by NH$_3$ and CH$_3$CN (Purcell et al. 2006),
a mm, sub-mm and far IR counterpart. In addition there is
evidence for molecular infall and outflow motions, as well as shocked 
gas traced by 4.5$\,\mu$m excess from Spitzer GLIMPSE data.

In both cases the H$_2$CO masers are found toward objects
that have no compact 6$\,$cm radio continuum emission
at the few mJy/beam level. 
Given the evidence of active massive star formation in
G23.01$-$0.41 and G25.83$-$0.18, the absence of radio 
continuum suggests that the regions are in a very early 
phase of massive star formation, prior to the development 
of radio-bright ultra-compact H{$\,$\small II} regions.
H$_2$CO 6$\,$cm masers appear to preferentially pinpoint very 
young massive stellar objects that may be categorized as 
HMPOs (High-Mass Protostellar objects, Beuther et al. 2007a) or, in the 
Zinnecker \& Yorke (2007) nomenclature, somewhere 
in the HDMC (Hot Dense Massive Core) or DAMS
(disk-accreting main-sequence star) phase.

\acknowledgments

E. A. is supported by a NRAO predoctoral fellowship. 
H. L. was supported by a postdoctoral stipend from the 
German Max Planck Society. 
L. O. was supported in part by the Puerto Rico 
Space Grant Consortium.
We thank the anonymous referee for comments that
improved the manuscript. 
This research made use of the NASA's Astrophysics Data System,
the Avedisova (2002) catalog,
the VizieR catalog access tool (CDS, Strasbourg, France),
and archival data from the {\it Spitzer Space Telescope}.
The {\it Two Micron All Sky Survey} catalog was also
consulted as part of this investigation; 2MASS is a joint project of the 
University of Massachusetts and the Infrared Processing and 
Analysis Center/California Institute of Technology, funded 
by NASA and the National Science Foundation.

\clearpage

\begin{deluxetable}{lc}
\tabletypesize{\scriptsize}
\tablecaption{VLA H$_2$CO Survey \label{tbl-1}}
\tablewidth{0pt}
\tablehead{
\colhead{Parameter}   & \colhead{Value} }
\startdata
\rm Date                   & 2006 Jun. 13 - 15                 \\
Configuration$^a$          & BnA$\rightarrow$B                 \\
$\nu_o\,$(GHz)$^b$         & 4829.6594                         \\
IF Mode                    & 2IF (AD)                          \\
BW~(MHz)$^c$               & 1.56                              \\
~~~~~~(\kms)               & 97.0                              \\
$\Delta\nu^d$ (kHz)        & 7.3                               \\
~~~~~~(\kms)               & 0.46                              \\
Flux Density Calib.        & 3C$\,$286                         \\
~~~~~~Assumed S$_\nu$ (Jy) & 7.52                              \\
Phase Calib.               & J1832$-$105                       \\
~~~~~~Measured S$_\nu$ (Jy)& 1.26                               \\
Phase Calib.               & J1824$+$107                       \\
~~~~~~Measured S$_\nu$ (Jy)& 0.80                               \\
Phase Calib.               & J1950$+$081                       \\
~~~~~~Measured S$_\nu$ (Jy)& 1.07                               \\
\enddata
\tablenotetext{a}{Reconfiguration from BnA to B array.}
\tablenotetext{b}{Weighted average frequency of the 
H$_2$CO J$_{K_aK_c} = 1_{10} - 1_{11}$ $F=2-2$ and $F=0-1$
hyperfine transitions given in Tucker et al. (1970).}
\tablenotetext{c}{Bandwidth per IF.}
\tablenotetext{d}{Spectral resolution.}
\end{deluxetable}

\clearpage

\begin{deluxetable}{lcccll}
\tabletypesize{\scriptsize}
\tablecaption{Observed Targets. \label{sources}}
\tablewidth{0pt}
\tablehead{
\colhead{Target} & \colhead{$\alpha$(2000)$^{\blacklozenge}$} 
& \colhead{$\delta$(2000)$^{\blacklozenge}$} &
\colhead{V$_{\mathrm{LSR}}^{\diamond}$} & Phase Calib. & rms$^{\ddagger}$ \\
\colhead{} & \colhead{(h m s)} & \colhead{(\arcdeg\phn\arcmin\phn\arcsec)} &
\colhead{(\kms)} & \colhead{} & \colhead{(mJy)} \\
}
\startdata
G13.13$-$0.15             & 18 14 41.27 &$-$17 37 04.2 & 40.0	& J1832$-$105 & 7.6 \\
G13.21$-$0.14             & $\dag$      & $\dag$       & 40.0	& J1832$-$105 & 17.4\\
G21.57$-$0.03             & 18 30 36.50 &$-$10 06 44.0 & 110.0	& J1832$-$105 & 5.7\\
G22.05$+$0.22             & 18 30 35.70 &$-$09 34 26.0 & 50.0	& J1832$-$105 & 5.2\\
G22.36$+$0.06             & 18 31 44.84 &$-$09 22 08.5 & 85.0	& J1832$-$105 & 4.9\\
G23.01$-$0.41             & 18 34 40.60 &$-$09 00 28.0 & 75.0	& J1832$-$105 & 6.2\\
G22.97$-$0.39             & ${\star}$   & ${\star}$    & 75.0	& J1832$-$105 & 5.4\\
G24.68$-$0.16             & 18 36 51.80 &$-$07 24 48.3 & 50.0	& J1832$-$105 & 4.6\\
G25.79$-$0.14             & 18 38 51.73 &$-$06 25 00.6 & 100.0	& J1832$-$105 & 4.9\\
G25.80$-$0.16             & $*$         &  $*$         & 100.0	& J1832$-$105 & 5.2\\
G25.83$-$0.18             & $*$         &  $*$         & 100.0	& J1832$-$105 & 6.7\\
G27.28$+$0.15             & 18 40 33.90 &$-$04 57 18.0 & 35.0	& J1832$-$105 & 5.6\\
G30.87$+$0.11             & 18 47 15.76 &$-$01 47 10.3 & 90.0	& J1832$-$105 & 10.1\\
G33.92$+$0.11             & 18 52 50.27 &$+$00 55 29.6 & 100.0	& J1824$+$107 & 4.8\\ 
G37.87$-$0.40             & 19 01 53.57 &$+$04 12 49.2 & 60.0	& J1824$+$107 & 5.6\\ 
G39.10$+$0.48             & 19 00 59.50 &$+$05 42 28.0 & 15.0	& J1824$+$107 & 5.9\\
G42.43$-$0.26             & 19 09 50.08 &$+$08 19 39.7 & 60.0	& J1950$+$081 & 3.6\\
G43.26$-$0.21             & 19 11 05.74 &$+$09 05 06.0 & 60.0	& J1950$+$081 & 9.3\\
\enddata
\tablecomments{
$^{(\blacklozenge)}$ Phase tracking center.
$^{(\diamond)}$ Center bandpass LSR velocity. 
$^{(\ddagger)}$ Single channel rms of the final H$_2$CO data cube.  
$^{(\dag)}$ G13.21$-$0.14 was included in the 
VLA primary beam of G13.13$-$0.15.
The reported rms was obtained toward the G13.21$-$0.14 position 
after correcting for primary beam attenuation.
$^{(\star)}$ G22.97$-$0.39 was included in the 
VLA primary beam of G23.01$-$0.41.
The reported rms was obtained toward 
the G22.97$-$0.39 position after correcting 
for primary beam attenuation.
$^{(*)}$ G25.83$-$0.18 and G25.80$-$0.16 were included in the 
VLA primary beam of G25.79$-$0.14.
The reported rms values were obtained toward
the G25.83$-$0.18 and G25.80$-$0.16 positions after correcting 
for primary beam attenuation.}
\end{deluxetable}

\clearpage

\begin{deluxetable}{l ccc cc}
\tablecaption{Line Parameters of H$_2$CO 6$\,$cm Masers}
\tabletypesize{\scriptsize}
\tablewidth{0pt}
\tablehead{
\colhead{Source} &
\colhead{$\alpha_{peak}$(J2000)} &
\colhead{$\delta_{peak}$(J2000)} &
\colhead{I$_{\nu, peak}$}&
\colhead{V$_{\mathrm{LSR}}$} &
\colhead{FWHM} \\
\colhead{} &
\colhead{(h m s)} &
\colhead{(\arcdeg\phn\arcmin\phn\arcsec)} &
\colhead{(mJy/b)} &
\colhead{(km s$^{-1}$)} &
\colhead{(km s$^{-1}$)}
}
\startdata
\multicolumn{6}{c}{\vspace*{0.1cm} VLA BnA $\rightarrow$ B (Low Spectral Resolution) Observations }\\
\hline
G23.01$-$0.41$^\dag$    & 18 34 40.294(0.004) & $-$09 00 38.26(0.08) & 48(6) & 73.5(0.4) & $<$1.0 \\
G25.83$-$0.18$^\ddagger$& 18 39 03.627(0.003) & $-$06 24 11.18(0.06) & 85(7) & 90.2(0.4) & $<$0.8 \\
                        &                     &                      & 45(7) & 91.4(0.1) &1.8(0.3)\\
\hline
\multicolumn{6}{c}{\vspace*{-0.2cm} } \\
\multicolumn{6}{c}{\vspace*{0.1cm} VLA-D (High Spectral Resolution) Observations } \\
\hline
G23.01$-$0.41$^*$       & 18 34 40.29(0.06)   & $-$09 00 37.7(1.3) & 53(9)  & 73.60(0.09) & 0.4(0.2)  \\
G25.83$-$0.18$^\star$   & 18 39 03.69(0.07)   & $-$06 24 10.4(0.8) & 102(9) & 90.21(0.02) & 0.29(0.04)\\
                        &                     &                    & 52(9)  & 91.64(0.08) & 0.9(0.2)  \\
                        &                     &                    & 44(9)  & 90.6(0.1)   & 0.7(0.3)  \\
\enddata
\tablecomments{
Unless indicated otherwise, we list the intensity of the peak
channel (the rms is listed as uncertainty), the LSR velocity of 
the peak channel (channel separation reported as uncertainty),
and the FWHM (two times the channel separation is reported as 
uncertainty). 
$^{(\dag)}$ $\theta_{syn} = 1$\asec$6\times 1$\asec$1$, P.A. = $-$7\arcdeg.
$^{\ddagger}$~$\theta_{syn} = 1$\asec$6\times 1$\asec$1$, P.A. = $-$18\arcdeg.
The line parameters of the weak component were derived from a Gaussian fit.
$^{(*)}$~$\theta_{syn} = 21$\arcsec$\times 12$\arcsec, P.A. = $-$13\arcdeg.
$^{(\star)}$~$\theta_{syn} = 23$\arcsec$\times 15$\arcsec, P.A. =
$-$8\arcdeg. The line parameters were obtained from a three component
Gaussian fit; the intensity and
velocity of the two peak channels are: (113\mjyb, 90.2\kms)
and (66\mjyb, 91.7\kms); see Figure~2.}
\end{deluxetable}

\clearpage

\begin{figure}
\figurenum{1}
\includegraphics{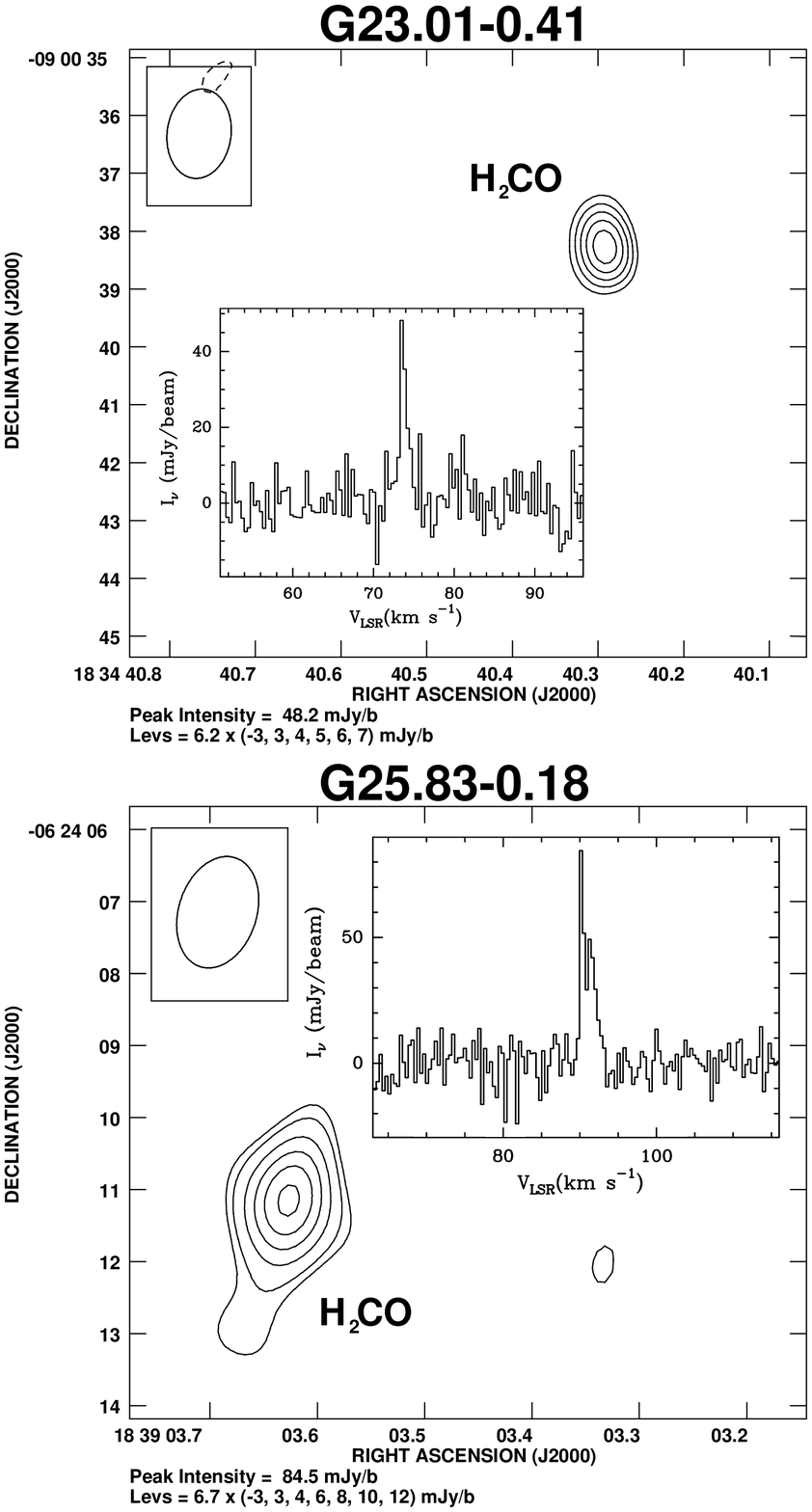} 
\vspace{18cm}\caption{Detection of H$_2$CO 6$\,$cm masers toward 
G23.01$-$0.41 (rms = 6.2\mjyb; 
$\theta_{syn} = 1$\asec$6\times1$\asec$1$, P.A. = $-$7\arcdeg)
and G25.83$-$0.18 (rms = 6.7\mjyb; 
$\theta_{syn} = 1$\asec$6\times1$\asec$1$, P.A. = $-$18\arcdeg). 
The peak channel images are
shown in contours and the spectra are shown in the insets.
The observations were conducted in 2006 June (see $\S2.1$), with
a spectral resolution of 0.46\kms.
\label{fig1}}
\end{figure}

\clearpage

\begin{figure}
\figurenum{2}
\includegraphics{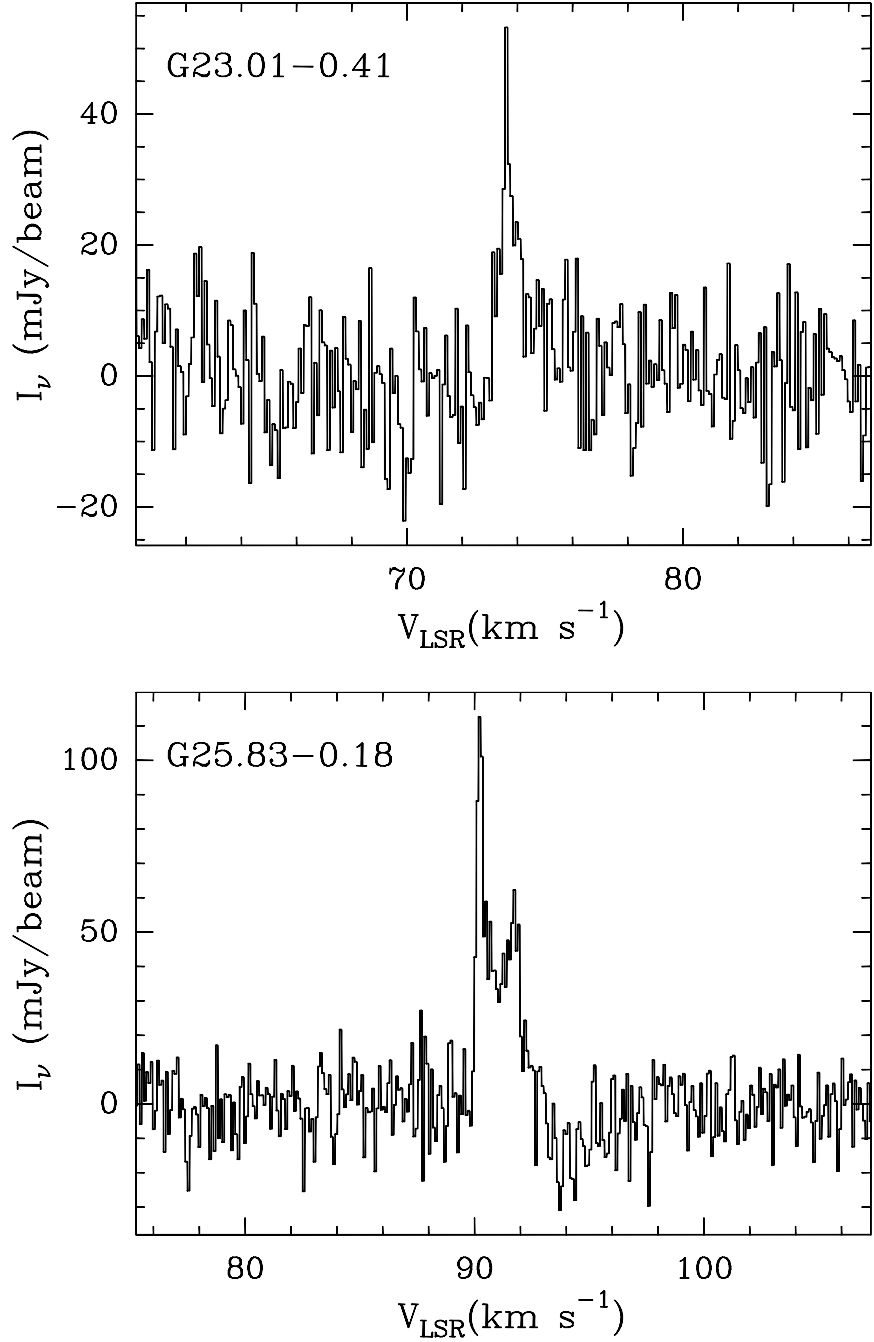} 
\vspace{18cm}\caption{High spectral resolution (0.11\kms)
VLA-D observations of the H$_2$CO 6$\,$cm masers in
G23.01$-$0.41 and G25.83$-$0.18. The G25.83$-$0.18 H$_2$CO spectrum
has a clear double peaked profile; the
line profile of the maser in G23.01$-$0.41 also appears to be the 
superposition of 2 lines. The observations were conducted in
2007 February ($\S 2.2$).
\label{fig2}}
\end{figure}

\clearpage
\begin{figure}
\figurenum{3}
\includegraphics{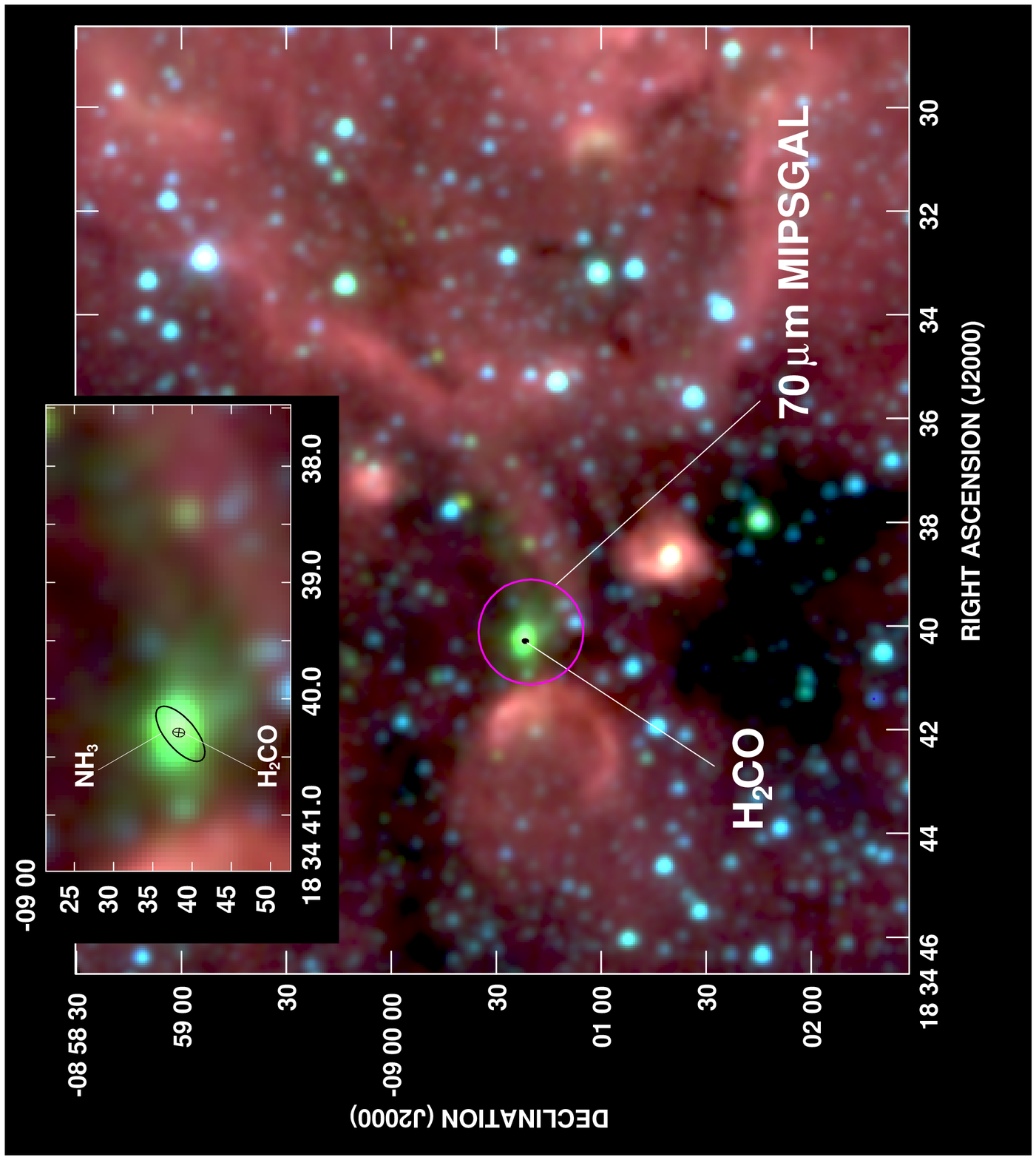} 
\includegraphics{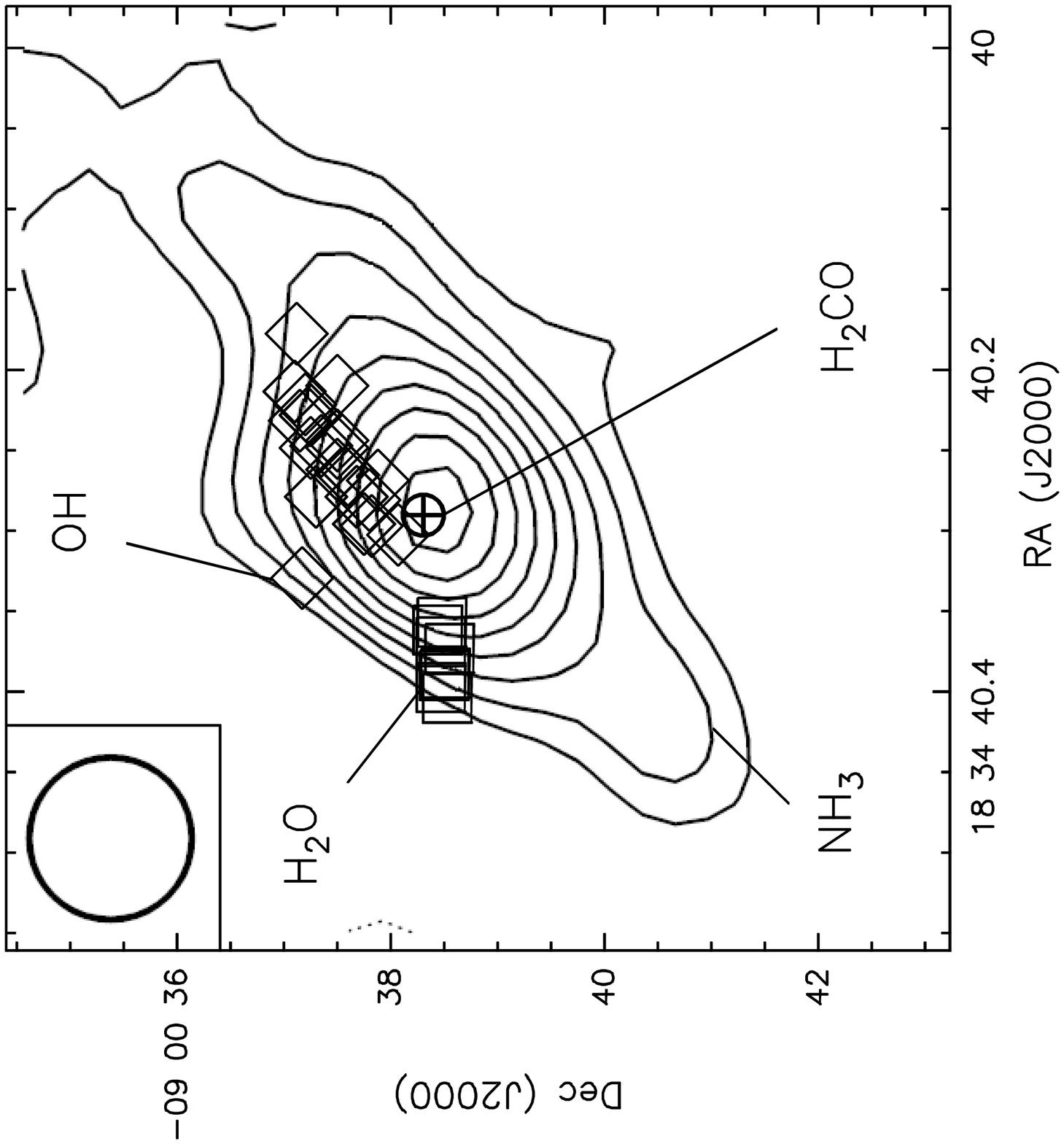} 
\vspace{14.4cm}\caption{
{\it Upper Panel:} Spitzer/IRAC GLIMPSE three-color image of the
G23.01$-$0.41 region; the 3.6, 4.5, and 8.0$\mu$m bands are shown
in blue, green, and red, respectively. The location and FWHM upper limit
of the 70$\,\mu$m MIPSGAL counterpart is shown with a magenta circle (see $\S 4.1.2$).
A closeup of the G23.01$-$0.41 region is shown in the inset. Note the 
strong 4.5$\,\mu$m excess toward the H$_2$CO maser that 
extends westwards. 
The ellipse represents the approximate size and position angle
of the NH$_3$ core shown in the lower panel (size, position and orientation
of the ellipse are from a fit of the 12$\,$\mjyb~contour 
of the Codella et al. (1997) map).
The location of the H$_2$CO maser is explicitly marked in all panels.
{\it Lower Panel:} Zoomed-in view of G23.01$-$0.41. The contours show
the NH$_3$ (3,3) emission imaged with the VLA by Codella et al. 
(1997; the original VLA image was convolved with a Gaussian profile
whose FWHM is shown in the upper-left corner of the figure).
The location of H$_2$O (22$\,$GHz) and OH (1665$\,$MHz) masers
from Forster \& Caswell (1999) are marked with squares and
diamonds, respectively. The absolute astrometry of the OH and
H$_2$O maser positions is $\sim 0$\asec 5 rms (see Forster \& Caswell 1999). 
\label{fig4}}
\end{figure}

\clearpage

\begin{figure}
\figurenum{4}
\includegraphics{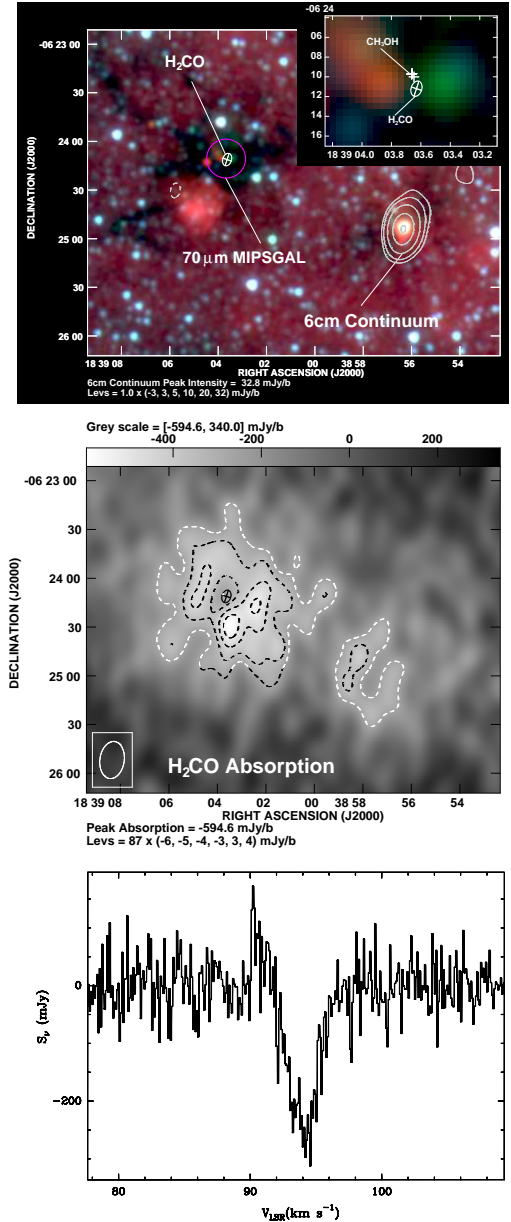} 
\vspace{15.5cm}\caption{{\it Upper panel:} We show in 
contours the radio continuum (6$\,$cm) image of
the G25.83$-$0.18 region (rms = 1\mjyb; 
$\theta_{syn} = 23$\asec$0\times15$\asec$0$, P.A. = $-$8\arcdeg)
superimposed on a Spitzer/IRAC GLIMPSE three-color image 
of the region (3.6$\mu$m blue, 4.5$\mu$m green, and 8.0$\mu$m
red). The location and FWHM upper limit of the 70$\,\mu$m 
counterpart is shown with a magenta circle (see $\S 4.2.2$).
A zoomed-in view of the H$_2$CO maser region is shown in the
inset; the position of CH$_3$OH 6.7$\,$GHz masers are
also shown (Walsh et al. 1998; absolute position accuracy $\sim 1$\arcsec).
{\it Middle Panel:} Sum of the channel maps in the 92.4 to 
95.5\kms~velocity range. 
The position of the H$_2$CO maser is shown with a `$\Earth$' symbol
in the upper and middle panels; the orientation of the 
symbol represents the P.A. of the $\theta_{syn}$ and its size
is five times the size of the $\theta_{syn}$ ($\S 2.1$, Table~3). 
In the inset of the upper panel, the size of the symbol
is that of the VLA synthesized beam.
{\it Lower Panel:} Spectrum of the G25.83$-$0.18
maser region obtained by integrating the brightness distribution
in a 85\arcsec$\times$90\arcsec~area centered at the 
H$_2$CO maser position. Note the maser emission at $\sim 90$\kms.
\label{fig4}}
\end{figure}

\end{document}